\documentclass[amsmath, amssymb, aps, prd, twocolumn, 
              ]{revtex4-2}

\usepackage{graphicx}
\usepackage{dcolumn}
\usepackage{bm}
\usepackage{longtable}
\usepackage{xcolor}
\usepackage[T1]{fontenc}

\begin{document}

\title{Studies of CPT symmetry in positronium decays with 192 plastic strip J-PET detector}

\author{N.~Chug$^{1,2}$}
\email{Contact email: neha.chug@uj.edu.pl}

\author{ S.~D.~Bass$^{1,3}$ }  
\author{ E.~Y.~Beyene$^{1,2}$ }
\author{ C.~Curceanu$^{4}$ } 
\author{ E.~Czerwi{\'n}ski$^{1,2}$ } 
\author{ M.~Das$^{1,2}$ } 
\author{ K.~V.~Eliyan$^{1,2}$ }
\author{ M.~Gorgol$^{5}$ }
\author{ J.~Hajduga$^{6}$ } 
\author{ S.~Jalali$^{1,2}$ } 
\author{ B.~Jasi{\'n}ska$^{5}$ }
\author{ K.~Kacprzak$^{1,2}$ }
\author{ T.~Kaplanoglu$^{1,2}$ } 
\author{ {\L{}}.~Kap{\l{}}on$^{1,2}$ } 
\author{ K.~Kasperska$^{1,2}$ }
\author{ A.~Khreptak$^{1,2}$ } 
\author{ A.~Kierys$^{7}$ }
\author{ G.~Korcyl$^{1,2}$ }    
\author{ T.~Kozik$^{1,2}$ }
\author{ K.~Kubat$^{1,2}$ } 
\author{ D.~Kumar$^{1,2}$ }     
\author{ A.~Kunimmal Venadan$^{1,2}$ } 
\author{ E.~Lisowski$^{8}$ } 
\author{ F.~Lisowski$^{8}$ } 
\author{ J.~M\k{e}drala-Sowa$^{1,2}$ } 
\author{ S.~Moyo$^{1,2}$ } 
\author{ W.~Mryka$^{1,2}$ }
\author{ S.~Nied{\'z}wiecki$^{1,2}$ } 
\author{ P.~Pandey$^{1,2}$ } 
\author{ S.~Parzych$^{1,2}$ } 
\author{ E.~P{\'e}rez~del~Río$^{1,2}$ } 
\author{ A.~Porcelli$^{1,9}$ } 
\author{ B.~Rachwa{\l{}}$^{6}$ } 
\author{ M.~R{\"a}dler$^{1,2}$ }
\author{ A.~Sienkiewicz$^{7}$ }
\author{ S.~Sharma$^{1,2}$ }
\author{ M.~Skurzok$^{1,2}$ } 
\author{ E.~{\L{}}.~St\k{e}pie{\'n}$^{1,2}$ }   
\author{ T.~Szumlak$^{6}$ } 
\author{ P.~Tanty$^{1,2}$ }
\author{ K.~Tayefi~Ardebili$^{1,2}$ } 
\author{ S.~Tiwari$^{1,2}$ } 
   
\author{P.~Moskal$^{1,2}$}%
\email{Contact email: p.moskal@uj.edu.pl}

\collaboration{J{-}PET Collaboration}

\affiliation{$^{1}$Faculty of Physics, Astronomy and Applied Computer Science, Jagiellonian University, Kraków, Poland}%

\affiliation{$^{2}$Centre for Theranostics, Jagiellonian University, Kraków, Poland}%

\affiliation{$^{3}$Kitzbühel Centre for Physics, Kitzbühel, Austria.}%

\affiliation{$^{4}$INFN, Laboratori Nazionali di Frascati, Frascati, Italy}%

\affiliation{$^{5}$Institute of Physics, Maria Curie-Sklodowska University, Lublin, Poland}%

\affiliation{$^{6}$AGH University of Cracow, Poland}%

\affiliation{$^{7}$Institute of Chemical Sciences, Faculty of Chemistry, Maria Curie-Sklodowska University, Lublin, Poland}%

\affiliation{$^{8}$ Cracow University of Technology, Faculty of Mechanical Engineering, Kraków, Poland}%

\affiliation{$^{9}$Centro de Investigaci\'on, Tecnolog\'ia, Educaci\'on y Vinculaci\'on Astron\'omica, Universidad de Antofagasta, Avenida Angamos 601, Antofagasta 1240000, Chile}%

\date{\today}

\begin{abstract}

A direct test of the CPT symmetry is performed for the electromagnetic decays of  ortho-positronium using the J-PET tomograph. We present the precise measurement of the CPT-sensitive angular correlation entailing the positronium spin and the momenta of its annihilation photons, surpassing previous studies utilizing the same detection system. Positrons originating from a $^{22}$Na source are emitted from the detector's center and subsequently form positronium atoms within the spherical chamber covered with porous material. Reconstruction of annihilation locations using the 192-strip J-PET detector makes it possible to determine the positronium emission direction, which defines the quantization axis along which positronium is polarized, without the application of external magnetic fields. The measurements were performed  in total for 356 days resulting in an identification of 47.8 million events with ortho-positronium decays into three photons. The results are consistent with the exactness of CPT symmetry with measured asymmetry amplitude 
-0.00029 $\pm$ 0.00022 (stat.) and  with statistical error four times smaller than the previous best measurement.  

\end{abstract}

\maketitle

\section{\label{sec:level1}Introduction}

CPT symmetry is one of the bedrocks of relativistic quantum field theories 
and, hence, has been the subject of extensive experimental tests.
Invariance under the combined operations of charge conjugation, C, parity transformation, P, and time reversal, T, is a fundamental property of local quantum field theories with a Hermitian Hamiltonian, invariance under proper Lorentz transformations and spin-statistics~\cite{Bjorken:1965zz,Feynman:1988wh}. 
CPT invariance holds independent of possible violations of individual discrete symmetries, e.g., the P and CP violation found within weak interactions.
Experimental tests have been conducted, e.g., with entangled kaons \cite{KLOE-2:2022hrb}, 
with neutrinos~\cite{Barenboim:2017ewj}, in  B-meson decays \cite{LHCb:2016vdl} and with antiprotons~\cite{Nowak:2024hvg, borchert202216, baker2025precision}. For single electrons an especially interesting test involves the electron anomalous magnetic moment $a_e = (g-2)/2$, with experimental value in excellent agreement with the QED prediction~\cite{Aoyama:2017uqe} 
to one part in $10^{12}$~\cite{Fan:2022eto}. 
Since QED respects CPT,  
this agreement of the electron $a_e$ measurement with QED theory 
is an implicit test of CPT. 
Any extra CPT violating interaction would change the theoretical prediction for $a_e$ so the measurement acts as a constraint on such possible interactions.
A more direct test comes from comparing measurements of electron and positron 
magnetic moment values.  
The result is consistent
with CPT symmetry holding
to ${\cal O}(10^{-12})$~\cite{VanDyck:1987ay}. 
If CPT were to fail, then (at least) one of the key input assumptions of a local quantum field theory, an Hermitian Hamiltonian, invariance under proper Lorentz transformations and spin statistics would also be failing.
Lorentz invariance~\cite{Kostelecky:2008ts} and 
spin-statistics~\cite{Baudis:2024vog} are so far 
working very well in all present experimental tests.
There are theoretical  ideas that any violation of Lorentz invariance might start only at 
${\cal O}(\Lambda_{\rm ew}^2/M^2)$ where 
$\Lambda_{\rm ew} \approx 246$~GeV 
is the electroweak scale and $M$ is the 
scale of ultraviolet completion 
(or upper-energy 
 limit of the theory)~\cite{Bjorken:2001pe}. 
For a possible 
emergent 
Standard Model 
this scale is about $10^{16}$ GeV 
\cite{Bass:2023ece}. A recent review of CPT violation scenarios is given in~\cite{lehnert2016cpt}.
 
Ortho-positronium, o-Ps, is special in that, 
as an unstable state, it is not an 
eigenstate of time reversal symmetry T or of CPT. 
It is an eigenstate of C and CP. 
Recent reviews of positronium physics are given in~\cite{Bass:2023dmv,Bass:2019ibo,Adkins:2022omi}.
In vacuum the o-Ps falls apart into three massless photons after a mean lifetime 
of 142 nanoseconds.
This has the  
interesting consequences that, 
while the underlying QED interactions are expected to preserve CPT, some observables of 
o-Ps decays can mimic CPT 
and CP 
violation through final state interactions (FSI)~\cite{bernreuther1988test,arbic1988angular}.
Experiments have focused so far on 
the CPT odd correlation
\begin{equation}\label{theta_def}
    {\rm O_{CPT}} = {\hat S}\cdot(\vec{k}_1 \times \vec{k}_2) / | \vec{k_{1}} \times \vec{k_{2}} | , 
\end{equation} 
where $\hat{S}$ is the unit spin vector of the spin-one o-Ps and 
$k_1 \geq k_2 \geq k_3$ 
denote the momenta of
the three decay photons numbered according to decreasing energy.
One expects a finite 
correlation value at the 
level of ${\cal O}(10^{-9}) - {\cal O}(10^{-10})$~\cite{bernreuther1988test}.
This CPT mimicking effect comes from 
FSI with the leading contribution 
coming from 
light by light scattering of two of 
the three photons in 
the final state. The experimental challenge is to search for and observe this effect. 

Recounting previous o-Ps tests of CPT symmetry, the first experimental test of the CPT sensitive angular correlation was performed by Arbic et al. in 1988~\cite{arbic1988angular}. These authors used an array of NaI detectors where the scintillators were arranged in a way to record a single annihilation plane of decaying o-Ps. A polarized positron beam was used to fix the Ps spin direction. They reversed the direction of the normal to decay plane and o-Ps spin and estimated the asymmetry ratio by averaging over the recorded events with two different spin directions. These studies were sensitive to geometrical asymmetries and found no CPT violation at 
the precision level of 0.014 $\pm$ 0.019. This approach of recording the up-down asymmetry due to the experimental construction marks an important difference from more recent tests with 
Gammasphere and 
with 
the Jagiellonian Positron Emission Tomograph (J-PET).

After Ref.~\cite{arbic1988angular}, 
Vetter and Freedman  
used 
the 4$\pi$ Gammasphere detector 
with arrays of high-purity germanium (HPGe)~\cite{vetter2003search}. 
The set-up enabled to estimate the different orientations
corresponding to an angle 
$\theta$
between the initially fixed o-Ps spin direction and 
the a vector normal to the decay plane of 
the o-Ps, where
\begin{equation} 
    \rm O_{CPT} = \cos \theta .
\end{equation}
The up-down asymmetry
was estimated for all possible orientations $\theta$ of the decay plane and resulted in no observation of a 
CPT violating asymmetry at the precision of 0.0026~$\pm~0.0031$.

The most recent measurements used the J-PET detector, 
which enables reconstruction of the o-Ps $\rightarrow$~3$\gamma$ annihilation place. 
The precision of this test improved the accuracy 
by a factor of three with a result consistent with no CPT violating asymmetry 
at the level of 0.00067
$\pm 0.00095$~\cite{moskal2021testing}. Similar level accuracy confirmation of CP symmetry in o-Ps decays and using decay photon polarization observables was shown in~\cite{Moskal:2024jfu}.
As described in Section II below, the detector 
is constructed
from plastic scintillators and 
has a better timing and angular resolution for recording o-Ps~$\to$~3$\gamma$~events compared to the Gammasphere detector. 
In this setup, the spin estimation for each o-Ps decay is applied without the use of an external magnetic field. Estimation of
the expectation value of the CPT-odd angular correlation (determined for the whole $\theta$-angle region) makes the J-PET approach  
different from previous experiments. 
Here we report the 
J-PET CPT result with a factor of four improvement in precision beyond this previous measurement.

\section{Experimental setup}
\begin{figure*}
\centering
   \includegraphics[width=0.82\textwidth]{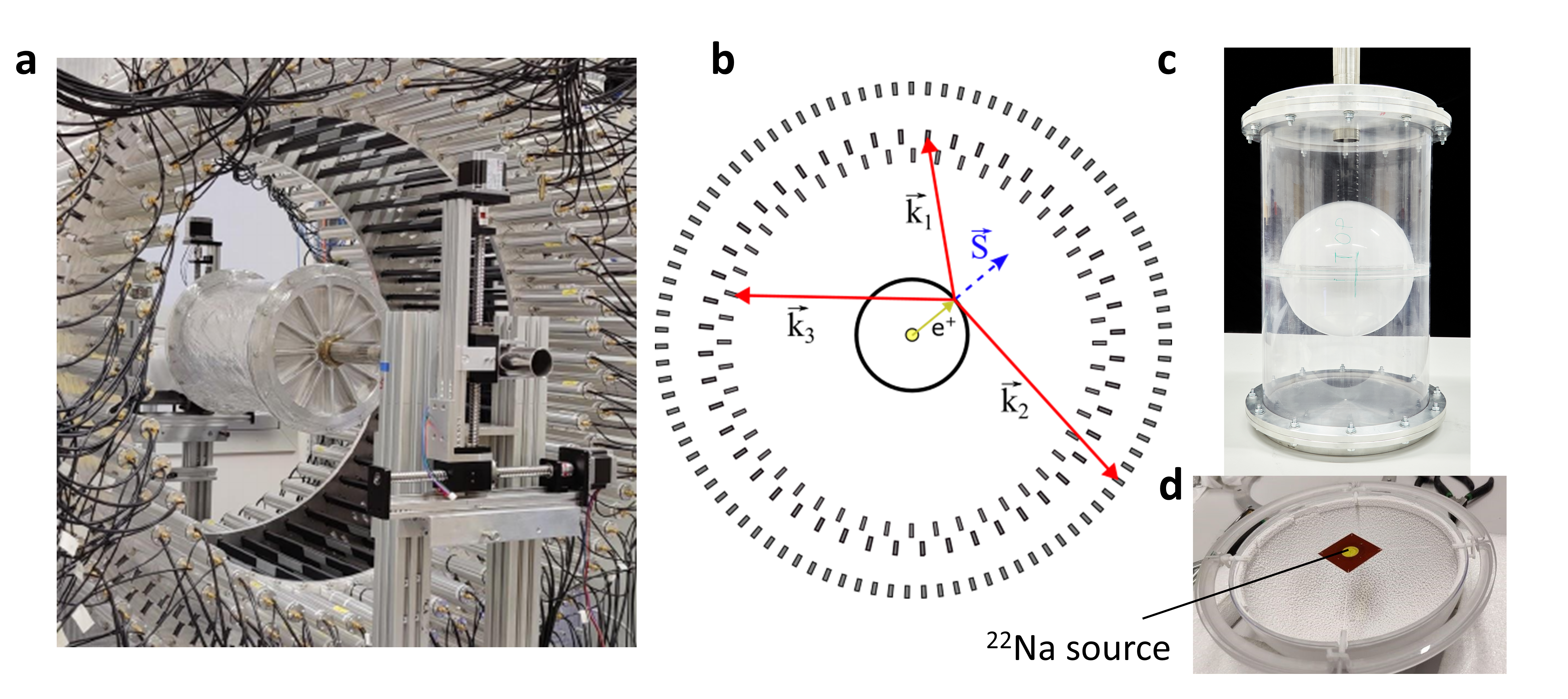}
   \caption{\label{fig:detector}\textbf{The experimental setup for CPT symmetry test.} \textbf{(a)} Photograph of the 192-strip plastic scintillator J-PET with an annihilation chamber in its center. \textbf{(b)} Schematic representation of the detector where three photons from o-Ps annihilation (arrows in red color) on the chamber's walls are emitted.  \textbf{(c)} Photograph of the positronium production medium used in the setup where the spherical annihilation chamber is placed inside a cylindrical tube enclosed by the endcaps at its two ends.  
   \textbf{(d)} The $^{22}$Na source sandwiched between Kapton foil (in yellow) is supported through strings and four bolt-like structures and placed on the source holder. This source holder is fixed at the center of the spherical chamber. The chamber setup is centered in the detector. 
   }  
\end{figure*}

The measurement of the CPT-sensitive correlation in o-Ps decays reported here
was conducted using the J-PET detector constructed from plastic scintillator strips~\cite{moskal2014test, niedzwiecki2017j, moskal2021positronium, moskal2025nonmaximal}. Fig.~\ref{fig:detector} illustrates the 
detector where 192 strips of plastic scintillators are arranged in a cylindrical configuration in three concentric layers~\cite{niedzwiecki2017j}. J-PET utilizes a triggerless data acquisition system (DAQ) comprising Field Programmable Arrays (FPGAs) implemented with Time-to-Digital Converters (TDCs) with a time resolution of 12~ps~\cite{korcyl2018evaluation}. Initially, the electrical signals from the Photomultiplier Tubes (PMTs) undergo probing at four distinct thresholds, constituting a multi-threshold system~\cite{palka2017multichannel, palka2014novel}. At each threshold level, the arrival time and width of the signal are observed and digitized via  TDCs. Consequently, four timing points are recorded for each signal at both the leading and trailing edges~\cite{sharma2020estimating}. Such data is collected from all PMTs within the detector and subsequently processed in the DAQ system using FPGAs, with a data rate of 80~MB/s.

The experimental setup constitutes the 3-layer J-PET detector system with an annihilation chamber for positronium production, which is presented in Fig.~\ref{fig:detector}(a) and (b). The chamber consists of two Plexiglas hemispheres with a radius of 10~cm each, shown in Fig.~\ref{fig:detector}(c). 
The inner walls of both hemispheres are coated with a 2~mm thick layer of mesoporous silica to enhance the positronium production. 
A positron source, $^{22}$Na, used in the measurement was prepared by evaporating
an aqueous solution of $^{22}$NaCl onto a 7.5$~\mu$m thick and 1.065~mg/cm$^2$ dense polyamide Kapton foil (shown in Fig.~\ref{fig:detector}(d)). The Kapton foil allows the transmission of around 92$\%$ of positrons emitted by $^{22}$Na isotope~\cite{gorgol2020construction, jasinska2016determination}. The source is placed along the equatorial plane of the hemisphere using a ring-like source holder made of plastic, as shown in Fig.~\ref{fig:detector}(d). 

The two hemispheres are joined to form a spherical annihilation chamber with the radioactive source at its center. 
The spherical chamber is enclosed in the center of a polycarbonate tube closed by two aluminum endcaps which serve as a vacuum vessel as shown in Fig.~\ref{fig:detector}(c). The outer tube is 43~cm long, and 3~mm thick with an inner radius of 12.2~cm. The small space between the spherical chamber and the walls of the cylindrical tube is kept for uniform pumping out of the air. Vacuum ($<$1~Pa) is maintained inside the chamber to minimize the scattering of positrons from the $^{22}$Na source~\cite{gorgol2020construction}. The whole chamber setup is placed inside the J-PET detector and is connected to the vacuum system through the long pipe at one endcap. The plane of the source holder is vertical, perpendicular to the axis of the cylindrical chamber. 

The measurement campaign for the CPT symmetry test with the J-PET detector and a spherical annihilation chamber took 1.3 years of data taking. During this period, three experiments were carried out with the same experimental setup except for different positron source activities (78 days with 1.1~MBq, 278 days with 4.0~MBq, and 60 days without the source). The set of multi-thresholds applied to the PMTs in these experiments are 30, 80, 190, and 300 mV~\cite{niedzwiecki2017j, sharma2020estimating}. The data from a total of 356 effective days of measurement with the positron emitting source were used for the CPT symmetry test, while a 60 days experiment without a source was used to estimated the background from cosmic radiation. The total volume of data collected from the above-mentioned measurements was around 2 petabytes. 
The measurements were conducted between April 2021 to August 2022.
In total, 47.8~million o-Ps signal events were identified.

\section{Event reconstruction and Signal selection}
\begin{figure*}
    \centering
    \includegraphics[width=0.94\textwidth]{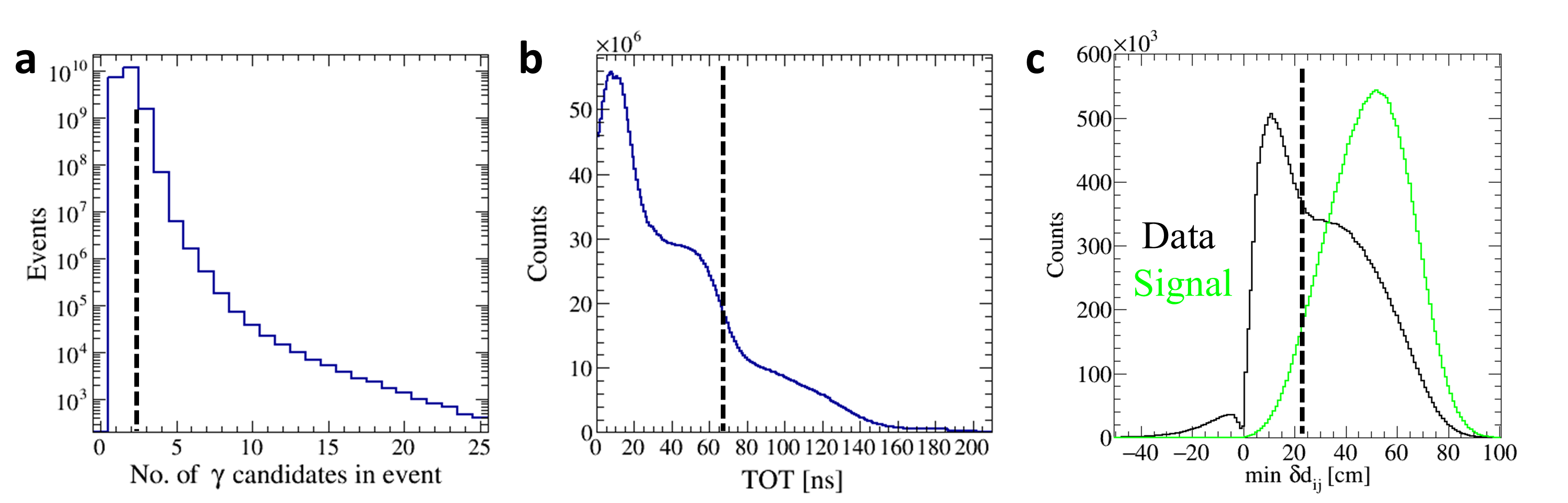}
    \caption{\label{fig:analysis_hit_based}\textbf{Hit-based analysis criteria for selection of o-Ps $\to$ 3$\gamma$.} \textbf{(a)} Distribution of a total number of $\gamma$ interactions in the detector referred to as hit multiplicity in an event. The events with a hit multiplicity greater than or equal to three are selected. \textbf{(b)} Time over Threshold (TOT) distribution for each hit in the selected event. The accepted region up to the Compton edge at 67 ns is for the annihilation events. \textbf{(c)} The smallest value of $\delta$d$_{ij}$ =  $ \mid$$\vec{r_{i}}$ - $\vec{r_{j}}\mid - c \mid \vec{t_{i}}$ - $\vec{t_{j}}\mid $ is plotted for each selected event. The distribution of min $\delta$d$_{ij}$ is compared for experimental data and o-Ps signal events from Monte-Carlo simulations. The hits with a greater time of interaction out of the hit pairs in the region $\delta$d$_{ij}$ $<$ 23 cm are discarded assuming those are the secondary Compton scatterings of the annihilation photons.  
    }    
\end{figure*}

The event selection criteria for this study were adapted from the previous work of testing CPT symmetry with J-PET~\cite{moskal2021testing}. The signal selection analysis was performed offline after collecting the data.  
The identification of o-Ps annihilation events within the detector relied on the selection of at least three annihilation photons interaction within a coincidence time window of 2.5~ns. The choice of time window is based on the detector geometry for the identification of 3$\gamma$ annihilation candidates. A topology of a typical signal event is shown in Fig.~\ref{fig:detector}(b). 
The distribution of the number of interactions (hits) in the scintillator for an event is shown in Fig.~\ref{fig:analysis_hit_based} (a), where events containing three or more hits are selected. The energy deposited by individual photons is estimated using the Time over Threshold (TOT) method~\cite{sharma2020estimating}. The technique proves valuable in differentiating prompt gamma (1275~keV) from annihilation photons originating from the decay of $^{22}$Na, due to the significantly higher average energy deposition of the prompt gamma. The experimental distribution of the TOT values for measurement with  $^{22}$Na source is shown in Fig.~\ref{fig:analysis_hit_based} (b). This distribution is composed of the Compton scattering spectrum, including contributions from 1275~keV de-excitation 
photon, 511~keV for back-to-back annihilation photons, photons with energy less than 511~keV originating from o-Ps decay, and photons from the secondary scattering within the detector. The Compton edge appears prominently for the 1275~keV 
photon and 511~keV photons at 130~ns and 67~ns, respectively. Since the energy of the photons from the o-Ps decay ranges from 0 to 511~keV, TOT values of 67~ns or below are used to primarily identify the annihilation photons. 

The secondary Compton scattering events constitute the one of the major background events in this study. Due to the sparse and segmented design of J-PET, these secondary interactions can be detected and distinguished from the primary annihilation signal. The scatter test based on $\delta d_{ij} = \lvert \Vec{r_{i}} - \Vec{r_{j}} \rvert - c \lvert t_{i} - t_{j} \rvert  $ is used to identify such events, where $c$ denotes the speed of light in vacuum, 
($t_{i}$, $\Vec{r_{i}}$) and ($t_{j}$, $\Vec{r_{j}}$) denote registered hit times and positions for the i$^{th}$ and j$^{th}$ hit. If the $i^\mathrm{th}$ and $j^\mathrm{th}$ hit were due to the same photon scattering, then $\delta d_{ij}$ would be equal to zero. 
$\delta d_{ij}$ is calculated for all possible hit pairs in an event containing three or more hits and the distribution of the minimum of the absolute values of $\delta d_{ij}$ is shown in Fig.~\ref{fig:analysis_hit_based}~(c). The reference cut value around the valley-like region at 23 cm is chosen to remove scattered hits in an event. 

The events with exactly three-hit multiplicity are considered for further processing. A three-dimensional annihilation point reconstruction is performed for the selected three hit events using the trilateration method~\cite{gajos2016trilateration}. The method enables to calculate the intersection point of three spheres centered on the hit positions of the three detected photons.  
The resolution of the reconstructed 3$\gamma$ vertex achieved with this method ($\approx$ 8~cm) is sensitive to the hit time resolution. 29$\%$ of the total selected three-hit events are reconstructed using trilateration reconstruction. The remaining 71$\%$ of events are rejected as the background. 

\begin{figure*}
    \centering
    \includegraphics[width=0.72\textwidth]{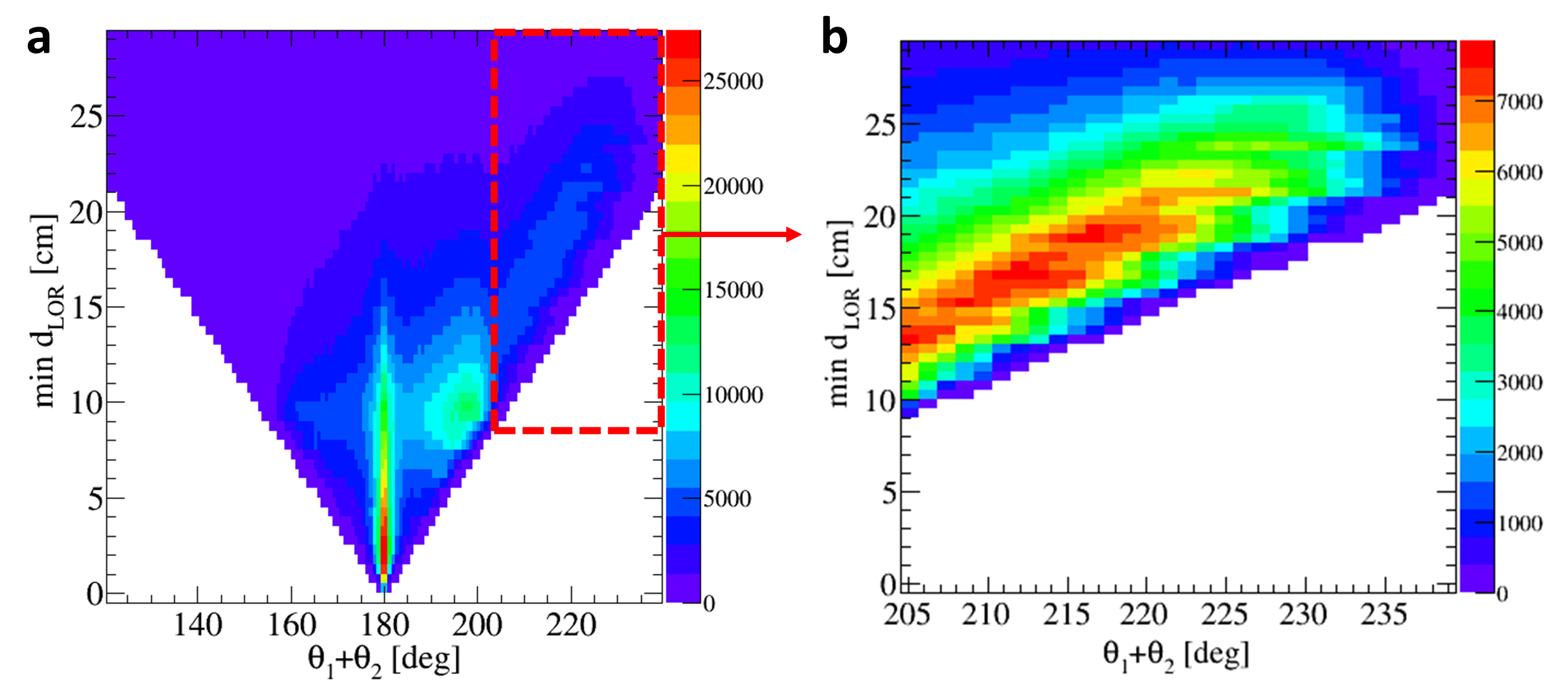}
    \caption{\label{fig:analysis_event_based}\textbf{Event-based analysis criteria for the selection of o-Ps $\to$ 3$\gamma$ events.} The events with a hit multiplicity equal to three are considered after the hit-based selection criteria. \textbf{(a)} The sum of the two smallest relative angles between the photons vs. the smallest distance between the hypothetical annihilation point of 2$\gamma$ on the Line of Response (LOR) and the center of detector. The angles $\theta$ are calculated assuming that photons originate from the center of the detector. The highly concentrated region around $\theta_{1}$ + $\theta_{2}$ $=$ 180$^{\circ}$ is the 2$\gamma$ annihilation events from the direct annihilation in the source. The other concentrated region around 200$^{\circ}$ is the contribution from events like p-Ps annihilation in the porous material at the wall of the annihilation chamber and secondary Compton scatterings of annihilation photons. \textbf{(b)} Events in the region $\theta_{1}$ + $\theta_{2}$ $>$ 204$^{\circ}$ are identified as originating from the o-Ps annihilation. }    
\end{figure*}

The significant amount of background events in this study also comes from the two back-to-back photons of energy 511 keV each from direct annihilation, accompanied by the prompt photon from the $^{22}$Na source. These annihilation photons have a collinear momentum and originate from the detector's center, where the source is placed. To identify such background events, the relative angles between photon momenta from the center of the detector system are considered, as they amount to roughly 180$^\circ$. 
Moreover, the distance from the 2$\gamma$ annihilation point on its Line of Response (LOR) to the source position would be comparatively smaller for these events compared to signal events. 
The 2$\gamma$ annihilation point is calculated based on hits position and time of flight (TOF) information~\cite{niedzwiecki2017j}.
The distribution for the sum of the two smallest relative angles ($\theta_{1}$ + $\theta_{2}$) and the minimum distance from the hypothetical 2$\gamma$ annihilation point to the source position (min d$_{LOR}$) is presented in Fig.~\ref{fig:analysis_event_based}~(a). Events from direct annihilation are anticipated to cluster in the region with a small min d$_{LOR}$ value and a sum of angles close to 180$^\circ$. The events from p-Ps annihilation at the surface of the spherical chamber are concentrated around the region of 200$^\circ$ angles. Events falling within this designated region $\theta_1 + \theta_2 > $~204$^\circ$ are identified as o-Ps events (Fig.~\ref{fig:analysis_event_based}~(b)), and are used for the CPT symmetry test. The cut value is determined from MC simulations by optimizing the signal-to-background ratio at different $\theta_1 + \theta_2$ values and a separate study of background events contributing to this study, explained more in the thesis prepared for this work~\cite{chug_thesis}.

\begin{figure*}
\centering
   \includegraphics[width=1\textwidth]{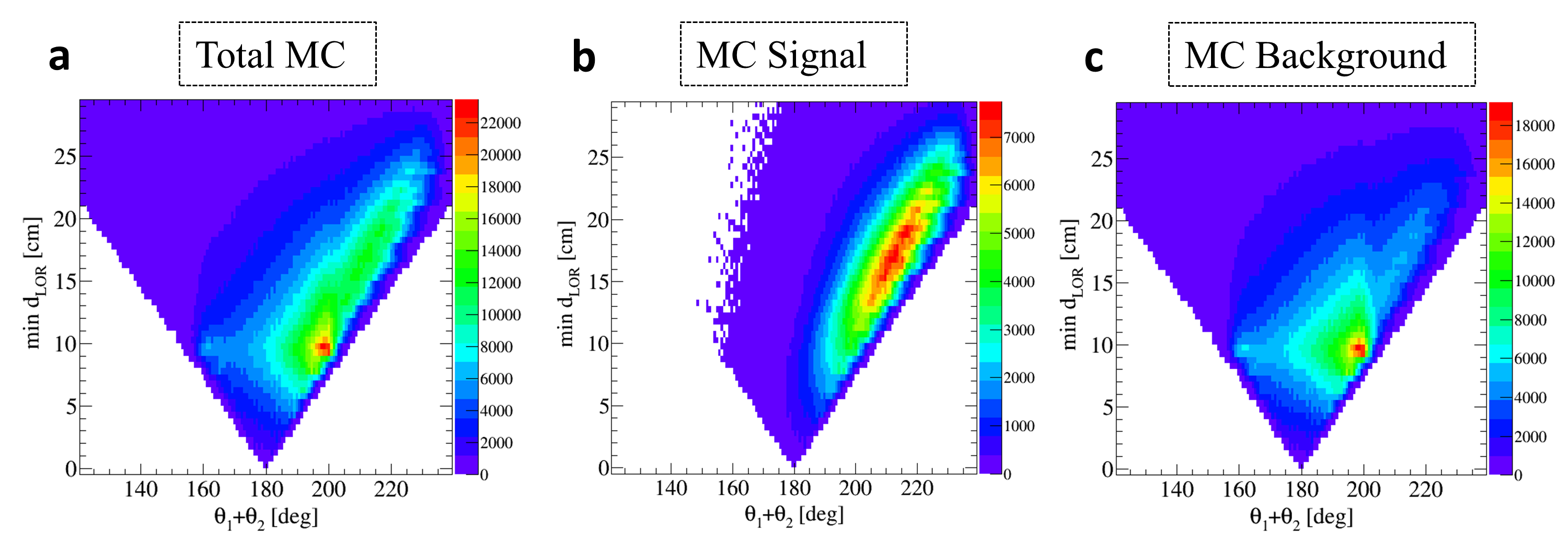}
   \caption{ \label{fig:angle_dlor} \textbf{Identification of o-Ps events from Monte-Carlo  simulations.} Distribution of sum of two smallest angles vs. smallest of distances from LOR is compared for \textbf{(a)} Total Monte-Carlo simulations that include the signal and different kinds of background in the study, \textbf{(b)} only signal events with three annihilation photons from o-Ps atom, and \textbf{(c)} three hit background events that can mimic the signal events in the study. The background event of 2$\gamma$ from direct annihilations is not shown here, therefore no structure is present around $\theta_{1}$ + $\theta_{2}$ $=$ 180$^{\circ}$ in (a) and (c) unlike in the experimental data Fig.~\ref{fig:analysis_event_based}a. Distribution (a), (b) and (c) are shown in different color scale to enhance the visibility of signal. }
   
\end{figure*}

Monte-Carlo simulations are used to help interpreting data at different stages of the analysis by comparing them with the experimental data. A similar distribution of sum of the two smallest relative angles between photons versus a minimum of d$_{LOR}$ are compared for the selected three-hit events using MC simulations, as shown in Fig.~\ref{fig:angle_dlor}. 

\section{Evaluation of CPT odd operator}

To reconstruct the angular correlation operator
in Eq.~(\ref{theta_def}), 
$ 
{\rm O_{CPT}} =
{\hat{S}}\cdot(\vec{k_{1}} \times \vec{k_{2}})/
| \vec{k_{1}} \times \vec{k_{2}} |, 
$ 
the spin vector $\hat{S}$ of the o-Ps is defined along the direction of flight of the positron due to the longitudinal polarization of positrons from $\beta^{+}$ decay~\cite{skalsey1985longitudinal}. It is taken as a unit vector from the positron source position to the annihilation point of o-Ps on an event-by-event basis. The momenta $\vec{k_{1}} $ and $\vec{k_{2}}$ of annihilation photons are estimated from the reconstructed hit positions of the
$3 \gamma$ interactions in the scintillator and the event’s reconstructed annihilation point~\cite{moskal2016potential}.
The determination of the level of potential CPT violation 
requires the estimation of its expectation value. 

The mean and the statistical error of the expectation value are calculated separately for each of the two measurements performed for 1.1~MBq and 4.0~MBq sources, as given in Table~\ref{tab:statistical_uncert}. 

The statistical uncertainty is scaled up to account for the amount of background expected to be present in the final data sample based on Monte-Carlo  simulations:

\begin{table*}
    \centering
    \caption{\label{tab:statistical_uncert} The identified o-Ps events and the statistical uncertainty on the expectation value of the CPT-odd operator $\langle $O$_{\text{CPT}}$$ \rangle$ for two measurements conducted with J-PET for the CPT symmetry test. The corrected value corresponds to the expectation value after accounting for background content from each measurement.} 
    \begin{ruledtabular}
    \setlength{\tabcolsep}{10pt}
    \renewcommand{\arraystretch}{1.3}
        \begin{tabular}{c c c c c}  
         Activity &  Measurement duration & Identified o-Ps events  &  Expectation Value  & Corrected value \\  
         \hline    
         1.1 MBq & 78 days & 2.8 $\times$ 10$^{6}$ & (-3.1 $\pm$ 2.5) $\times$ 10$^{-4}$  & (-3.1 $\pm$ 3.4) $\times$ 10$^{-4}$ \\        
         4 MBq & 278 days & 45 $\times$ 10$^{6}$ & (-0.98 $\pm$ 0.62) $\times$ 10$^{-4}$  & (-0.98 $\pm$ 0.85) $\times$ 10$^{-4}$ \\       
    
        \end{tabular}
    \end{ruledtabular}    
\end{table*}

\begin{equation} \label{error_scaled}
    \begin{gathered} 
    \sigma_{\rm signal}^{2} = \sigma_{\rm experiment}^{2} \cdot \frac{N_{\rm experiment}}{N_{\rm experiment} - N_{\rm background}} \, , \\
    N_{\rm background} = f \cdot N_{\rm experiment} \,.
    \end{gathered}
\end{equation}
where $f$ is the fraction of background present in each measurement. It is taken as a ratio of amount of background to the total MC simulated events (including signal and background) after the signal selection criteria.  
The weighted average of the expectation values from the two measurements is evaluated using the inverse variance weight method, which yields the result (-1.10~$\pm$~0.82)~$\times$~10$^{-4}$.

\section{Systematic effects}

\begin{figure}[h!]
\centering
    \includegraphics[width=0.40\textwidth]{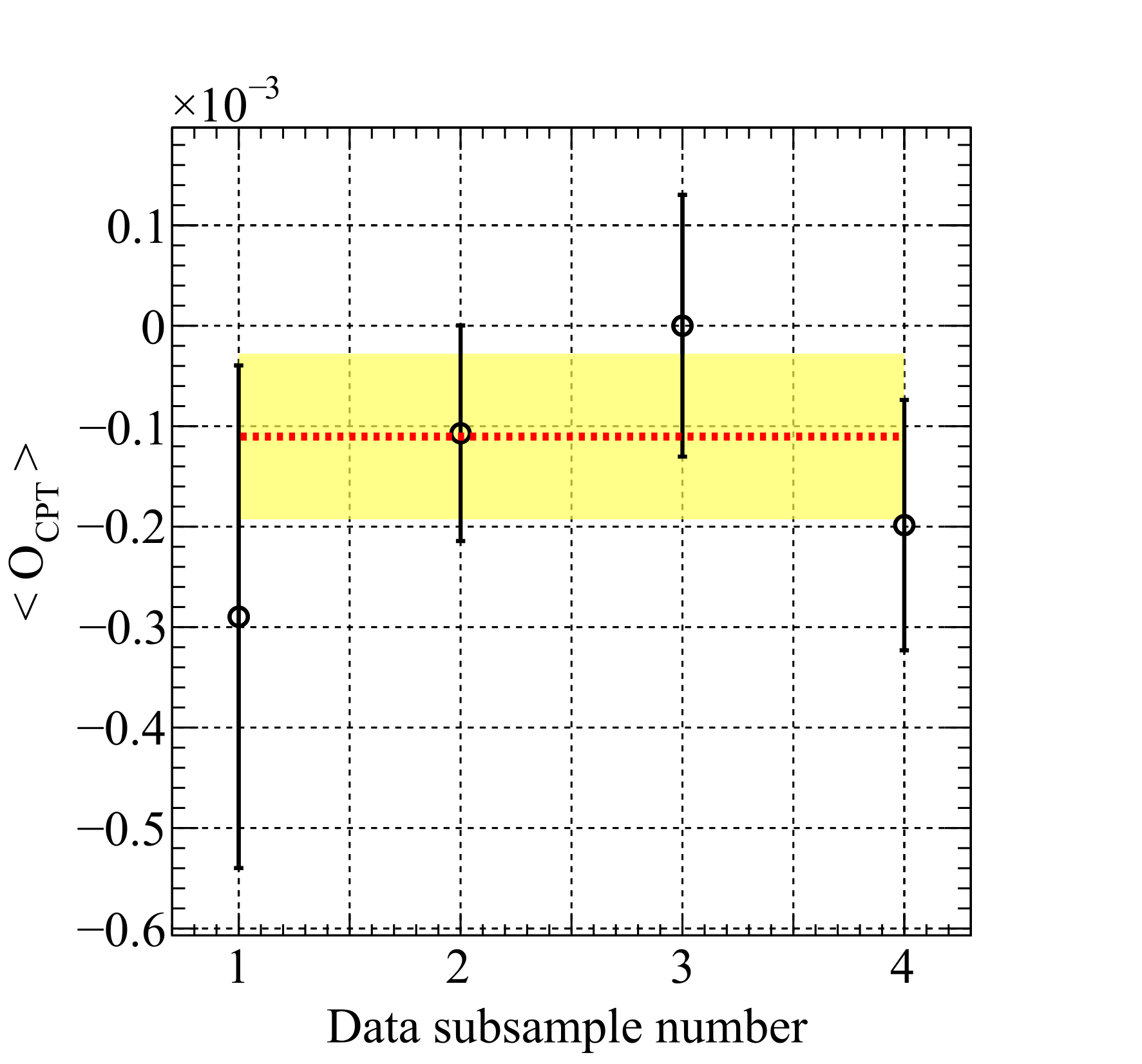}
   \caption{\label{fig:data_mix}\textbf{Data consistency check.} One year of experimental data is divided into four independent sub-samples.
    The single point represents the mean with its statistical error for the CPT
    odd operator for a sub-sample consisting of three months of data. The red and yellow lines represent the final mean and its error of $\langle$O$_{\text{CPT}}$$\rangle$ for the whole data sample.
  }  
\end{figure}

The different forms of systematic variation that could originate from the detector setup, background, and event selection criteria are checked.
The consistency of the data collected for the CPT symmetry test was cross-checked by splitting it into four independent sub-samples. Each sub-sample was generated by dividing one year of data into four parts, each consisting of three months of data. 
These sub-samples were analyzed separately to estimate the $\langle$O$_{\text{CPT}}$$\rangle$ of the CPT odd angular correlation operator for each sub-sample. The obtained values of $\langle$O$_{\text{CPT}}$$\rangle$ are consistent with the mean of the whole data observed as each sample point lies within 1$\sigma$ of the final results, as shown in Fig~\ref{fig:data_mix}.

To check 
for any asymmetry from the detector setup, a test was done where the signals from one of the scintillators are excluded from the analysis. 
This is a test of the extreme scenario assuming that we consider the detector as working properly and it is not working at all. The three annihilation photon events from 
the o-Ps decay are identified using the same analysis approach and the distribution of $\cos \rm \theta$ for the 
CPT odd angular correlation is studied.  
The expectation value of the CPT odd operator with one excluded scintillator in the detector comes out to be $\langle$O$_{\text{CPT}}$$\rangle$~=~(-3.1~$\pm$~2.5)~$\times$~10$^{-4}$ and (-1.1 $\pm$~0.7)~$\times$~10$^{-4}$ for 78-days and 278-days measurement with 1.1~MBq and 4 MBq source activity respectively and the errors are statistical.
There is no observed asymmetry on the final distribution if any scintillator in the J-PET detector is missing or stops working. 
The possible worse performance of any of the scintillator strips is impacting only the statistics of events.

One of the backgrounds in the study is cosmic radiation. Its contribution to the systematic uncertainty is estimated by analyzing the 60-day data from cosmic measurements. It is estimated that less than 0.1$\%$ of background originates from cosmic radiations in the final data sample from 356-day measurement with the radioactive source. 
The expectation value of 4123 cosmic events from cosmic measurements is $\langle$O$_{\text{CPT}}$$\rangle$~=~(2.1~$\pm$~6.2)~$\times$~10$^{-3}$ Therefore the maximal systematic uncertainty of the final result due to the presence of 0.1$\%$ cosmic events in the final event sample can be neglected at the level of 10$^{-5}$~\cite{chug_thesis, moskal2021testing}.

Any asymmetric effect due to the nonuniform thickness of the porous silica used in the annihilation chamber was checked. 
The azimuthal angle of the reconstructed annihilation points of the identified o-Ps to 3$\gamma$ decays in the XY plane is used to check the uniformity of porous material. The distribution should be uniform in case of the uniform thickness of coating in the experiment, while in the experimental data it is coming to be non-uniform. To check its effect on the final operator distribution, the Monte-Carlo simulations are re-weighted based on the scaling factors. 
There was no effect of non-uniform thickness of porous material in the final CPT odd operator distribution.

\begin{figure*}[t!]
\centering
   \includegraphics[width= 0.85\textwidth]{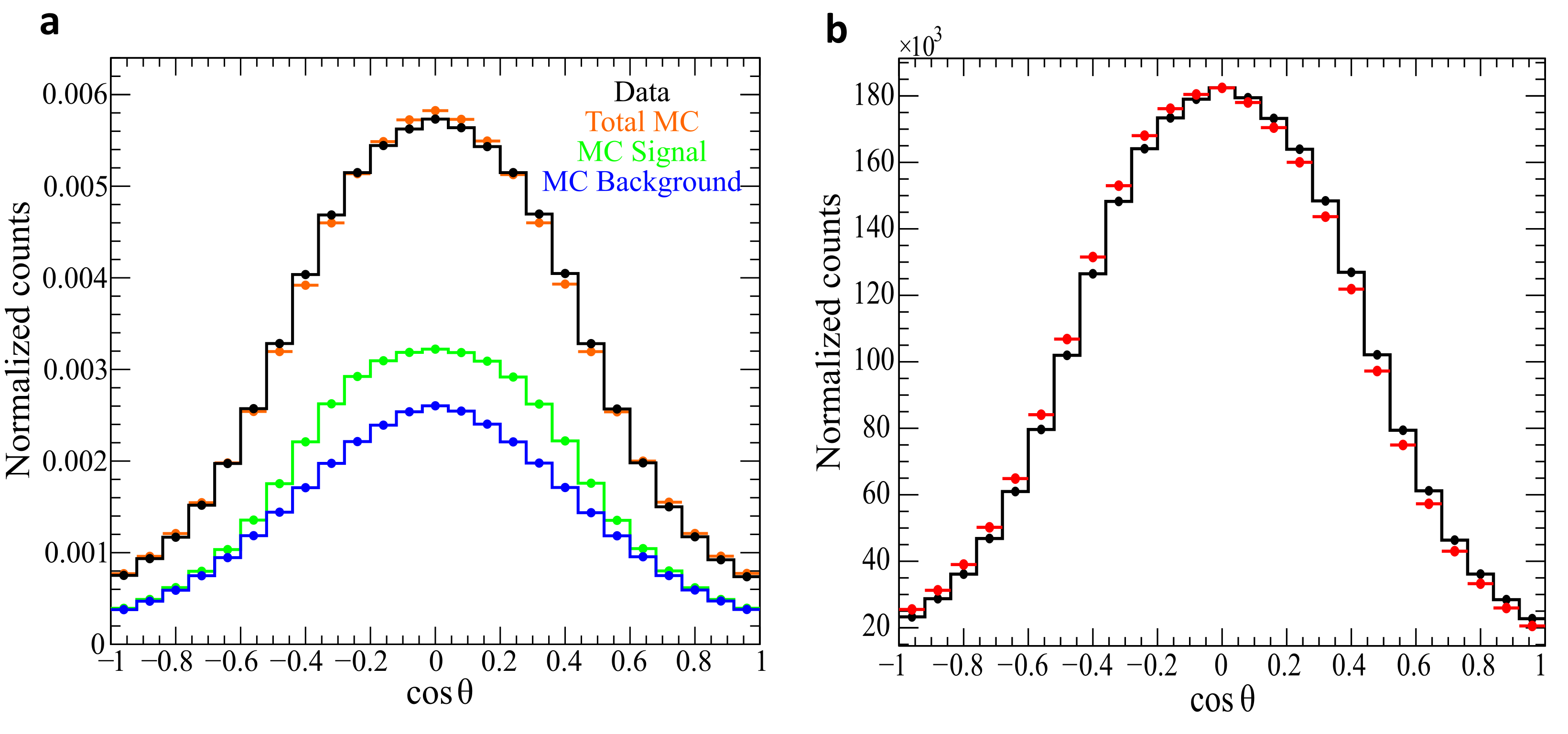}
   \caption{\label{fig:operator}\textbf{The CPT violating angular correlation operator ${\rm O_{CPT}}$.} 
   The distribution of the cosine of the angle between the reconstructed spin of o-Ps and normal to the orientation of its decay plane for the identified 2.8 million o-Ps events in experimental data (black) from 1.1 MBq source activity measurement. (a) The experimental data is compared to its corresponding total MC simulations (orange), MC simulated signal (green), and background (blue).  
    In (b), the red distribution represents artificially introduced asymmetry in data with asymmetry in the order of 10$^{-1}$.   
  }   
\end{figure*}

The impact of the event selection criteria (``cut values") on the expectation value of the CPT odd angular correlation is studied for the experimental data by varying the cut values within 
two  
standard deviations 
of their experimental resolution. There is no statistically significant impact on the measured value of $\langle$O$_{\text{CPT}}$$\rangle$ observed~\cite{chug_thesis}. 

The impact of the positioning of the $^{22}$Na source used in measurements on the experimental result is also evaluated. The source was positioned at the center of the detector system during the experimental measurements with the precision of the fraction of a millimeter. To study this effect, the coordinate system of reconstructed hits from the already existing data (with the source positioned at the center of the detector) is shifted from the center to a few other positions within 1$\sigma$ (1~mm) from the center. The result was not sensitive to these changes within the obtained statistical uncertainty.

\section{Results}

The expectation value of the CPT asymmetric operator estimated for the identified 
$47.8 \times 10^6$ o-Ps events, is equal to
\begin{equation}
  \langle {\rm O_{CPT}} \rangle ~=~ (-1.10 \pm 0.82 \ ({\rm stat.})) \times 10^{-4} .  
\end{equation}
The obtained value is the weighted average of expectation value of two measurements at 1.1 and 4.0~MBq source activity. 
The statistical error of expectation value for each measurement is scaled up for the background contribution as defined in Eq.~(\ref{error_scaled}).    
Based on a Monte-Carlo simulation study, the final samples for the two measurements consist of 55$\%$ and 53$\%$ signal at 1.1~MBq and 4~MBq, respectively. 
After correcting the expectation value with the analyzing power of the setup (where $P= 37.4\%$ is
the estimated polarization degree~\cite{moskal2021testing}), 
the asymmetry amplitude 
for any CPT violating angular correlation in o-Ps decays 
comes out to be 
\begin{equation}
    C_{\rm CPT} = \langle {\rm O_{CPT}} \rangle/P = -0.00029 \pm 0.00022 \ ({\rm stat.}) .
\end{equation}
The correction factor $P$ is the degree of polarization associated with the uncertainty in estimating the spin axis of ortho-positronium events.
It is not directly measured but takes into account the factors like average polarization of positrons from $^{22}$Na decay($\approx$ 67$\%$)~\cite{vetter2003search}, depolarization of positrons in porous material (8$\%$)~\cite{yang1997study}, ortho-positronium polarization due to spin statistics, $P_{\rm o-Ps} = \frac{2}{3} P_{e^{+}}$~\cite{arbic1988angular}, and polarization loss due to geometrical uncertainty ($\approx$ 9$\%$)~\cite{gajos2016trilateration}.
The final obtained results are consistent with the CPT invariance showing no asymmetry at the achieved level.

The distribution of the angular correlation (defined in Eq.~(\ref{theta_def})) is shown in Fig.~\ref{fig:operator} (a) for one of the experimental runs with J-PET and its corresponding MC simulations. 
In order to understand how a CPT asymmetry would influence the distribution of the angular correlation, we add here a red histogram (in Fig.~\ref{fig:operator} (b)) showing the distribution for an artificially introduced asymmetry at the level of 10$^{-1}$ (right side figure). It is implemented on the experimental data (black distribution) using a model probability function $Prob({\rm O^{'}_{CPT}}) = (a*\cos \theta + 1)\cdot Prob ({\rm O_{CPT}}) $ where the parameter $a$ is set to 0.1.

\section{Discussion}

\begin{figure}[t!]
    \centering
    \includegraphics[width=0.46\textwidth]{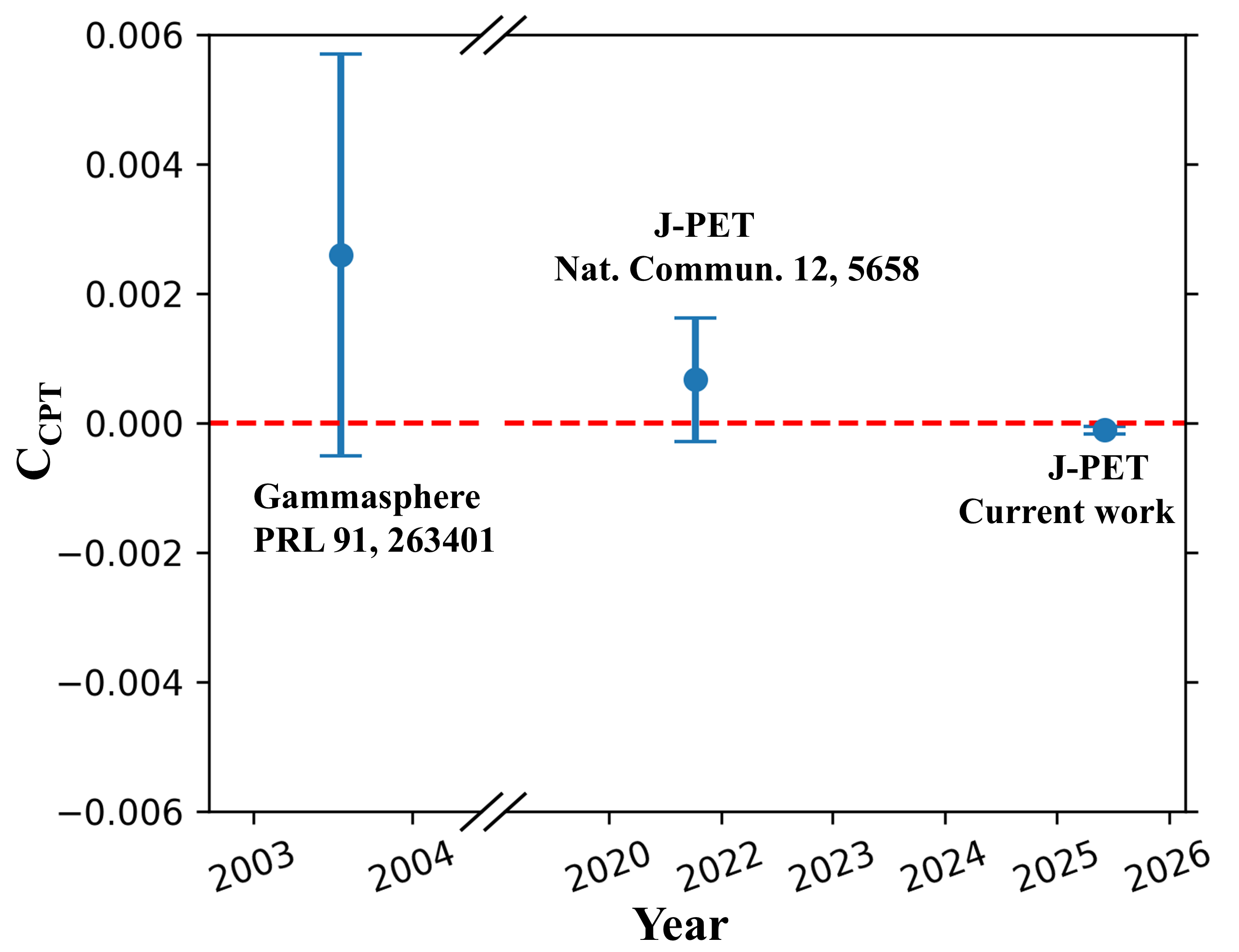}
    \caption{\label{fig:cpt_over_years}\textbf{Comparison of the precision CPT symmetry tests in 
    o-Ps~$\to 3 \gamma$ decays.} The amplitude of CPT sensitive angular correlation, $C_{\rm CPT}$, is shown for the three different experiments. The red horizontal line at zero represents no CPT symmetry violation. The error bars around each value correspond to the total uncertainty reported by each experiment.  }    
\end{figure}

Our result, Eq.~(5),  is 
the present most precise test of CPT symmetry in 
o-Ps decays.
The measurement was performed using 
the J-PET detector
by measuring the angular correlation between 
the plane spanned 
by the momenta of annihilation photons and the spin orientation of decaying positronium without using an external magnetic field. 
Fig.~\ref{fig:cpt_over_years} shows the obtained result compared with the previous two most accurate studies with the Gammasphere and J-PET detectors,
$C_{\rm CPT} = 0.0026 \pm  0.0031$~\cite{vetter2003search} and 0.00067 $\pm$ 0.00095~\cite{moskal2021testing}, respectively. The determined amplitude of the CPT symmetry violating correlations is consistent with zero within the achieved precision of 0.00022, a factor of four more precise than the last most accurate experiment~\cite{moskal2021testing}.

Our present measurement is consistent with the QED prediction that CPT symmetry is expected to hold in the o-Ps decay process.
Future measurements with the next generation of the
J-PET detector~\cite{ardebili2024assessing, tayefi2023evaluation, moskal2024first, das2024} will be continued.
The next generation high sensitivity total-body J-PET scanner under construction at the Jagiellonian University in Cracow~\cite{moskal2020prospects, moskal2021simulating, moskal2024vision}
should allow us to reach sensitivities on CPT-odd correlations with at least an order of magnitude improvement on the present measurement. 
These future measurements will approach much closer to the limit where CPT mimicking FSI effects~\cite{bernreuther1988test,arbic1988angular} are expected. 
As a  future project it would be interesting also to understand how these CPT measurements in o-Ps decays translate into constraints on CPT parameters in the Standard Model Extension effective theory framework~\cite{Colladay:1996iz}.

\section*{Acknowledgments}
We acknowledge support from the National Science Centre of Poland through grants 
2019/35/B/ST2/03562 (P.M.), 2021/42/A/ST2/00423 (P.M.), 2022/45/N/ST2/04084 (N.C.), 2023/50/E/ST2/00574 (S.S.), and 2020/38/E/ST2/00112 (E.P. del R.); the SciMat and qLife Priority Research Areas budget under the program Excellence Initiative - Research University at the Jagiellonian University (P.M. and E.Ł.S). We also acknowledge Polish high-performance computing infrastructure PLGrid (HPC Center: ACK Cyfronet AGH) for providing computer facilities and support within computational grant no.~PLG/2024/017688.  

We thank A.~Gajos for extensive discussions and general help in the initial phase of performing data analysis. We acknowledge the technical support of A.~Heczko, M.~Kajetanowicz, Dr.~P.~Kapusta, W.~Migdał, and A.~Mucha.

\end{document}